\documentclass[journal]{IEEEtran}
\usepackage{graphicx}

\begin{document}

\title{Outage Rates and Outage Durations of  \\ Opportunistic Relaying Systems }

\author{Zoran Hadzi-Velkov, and Nikola Zlatanov 
%..........
%\author{Zoran Hadzi-Velkov, Nikola Zlatanov

\thanks{Accepted for IEEE CommunLetters}
\thanks{Z. Hadzi-Velkov and N. Zlatanov are with the Faculty of
Electrical Engineering and Information Technologies, Ss. Cyril and
Methodius University, Karpos 2 bb, 1000 Skopje, R. Macedonia,
Email: zoranhv@feit.ukim.edu.mk, nzlatanov@manu.edu.mk}
%\thanks{G. K. Karagiannidis is with the Department of Electrical and Computer Engineering, Aristotle University of Thessaloniki, Thessaloniki, Email: %geokarag@auth.gr}
%\thanks{Digital Object Identifier 10}
\vspace{-10mm} }

\markboth{   }{Shell
\MakeLowercase{\textit{et al.}}: Bare Demo of IEEEtran.cls for
Journals} \maketitle

\begin{abstract}
Opportunistic relaying is a simple yet efficient cooperation
scheme that achieves full diversity and preserves the spectral
efficiency among the spatially distributed stations. However, the
stations' mobility causes temporal correlation of the system's
capacity outage events, which gives rise to its important
second-order outage statistical parameters, such as the average
outage rate (AOR) and the average outage duration (AOD). This
letter presents exact analytical expressions for the AOR and the
AOD of an opportunistic relaying system, which employs a mobile
source and a mobile destination (without a direct path), and an
arbitrary number of (fixed-gain amplify-and-forward or
decode-and-forward) mobile relays in Rayleigh fading environment.
\end{abstract}
\vspace{-0.1cm}
\begin{keywords}
Average outage rate, Average outage duration, Opportunistic
relaying, Doppler effect, Rayleigh fading \vspace{-0.2cm}
\end{keywords}
\vspace{-0.1cm}
\section{Introduction}\vspace{-0.1cm}
Cooperative diversity is a highly promising technique for coverage
extension and reliability improvement of wireless networks. It
exploits the additional degrees of freedom of the fading
environment, which is introduced by the spatially distributed
multiple relays utilizing either amplify-and-forward (AF) or
decode-and-forward (DF) relaying. Proposed recently in \cite{1},
the opportunistic relaying is a simple yet efficient cooperative
diversity protocol, whose diversity-multiplexing tradeoff is
identical to that of the more complex distributed space-time
coding cooperative schemes \cite{2}. By selecting a single ``best"
relay among the all available relays, the opportunistic relaying
achieves full spatial diversity while maintaining the spectral
efficiency of a two-hop communication link.

The outage and error probabilities of the opportunistic relaying
systems have been studied in \cite{3}-\cite{5}, which clearly
demonstrate its excellent performances. However, there are some
design issues for which the outage and error probabilities'
criteria are not sufficient, such as, packet or slot lengths and
latencies, switching rates \cite{5a}, power and bandwidth
allocation or, decision criterion for changing adaptive modulation
levels.

These issues can be addressed by investigating the system's
second-order outage statistics. To the best of authors' knowledge,
such statistics that describe the outage events of cooperative
systems, such as, the average outage rate (AOR) and average outage
duration (AOD), have not been studied previously. We propose the
AOR and the AOD be defined with respect to the capacity outage
events derived from information-theoretic capacity of the
opportunistic system. Similar definition of the outage statistics
has been applied over MIMO systems in \cite{5b}. In this letter,
we derive exact expressions for the AOR and the AOD of
opportunistic systems, employing either AF or DF relaying in
Rayleigh fading environment.

\section{Opportunistic relaying with mobile stations }
\subsection{System model}
\vspace{-0.0cm} Similarly to \cite{1}, we consider a typical
half-duplex dual-hop communication scenario, where the
communication between the source $S$ and the destination $D$ is
possible only via $M$ relays (denoted by $R_k$, $1\leq k\leq M$),
as the direct path is assumed blocked by an intermediate wall. In
the beginning of each slot (divided into two equal sub-slots), a
single ``best" opportunistic relay is selected out of the $M$
possible dual-hop paths for relaying the communication between $S$
and $D$. During the first sub-slot, $S$ transmits its signal over
the first hop, while the selected relay forwards that signal
toward $D$ over the second hop during the second sub-slot.

The channel is exposed to Rayleigh fading and is assumed to remain
constant during the entire slot duration. Without loss in
generality, we assume that $S$ and the selected ``best" relay
transmit with equal powers $P_T$, rendering the total available
transmission power to $2P_T$. Denoting the Rayleigh-faded channel
gains of hops $S\rightarrow R_k$ and $R_k\rightarrow D$ during a
given slot $t$ by $\alpha_{Sk} (t)$ and $\alpha_{kD}(t)$, the
received signal-to-noise ratios (SNRs) at the relay $R_k$ and at
the destination $D$ are expressed as $\gamma_{Sk} = P_T
\alpha_{Sk}^2(t) /N_0$ and $\gamma_{kD} = P_T \alpha_{kD}^2(t)
/N_0$, with $N_0$ as the noise power. Specifying the average
squared channel gains as $E[\alpha_{Sk}^2] = \Omega_{Sk}$ and
$E[\alpha_{kD}^2] = \Omega_{kD}$, the average received SNRs at
$R_k$ and $D$ are respectively given by $\bar \gamma_{Sk} = P_T
\Omega_{Sk} /N_0$ and $\bar \gamma_{kD} = P_T \Omega_{kD} /N_0$.

The relay selection is based on the estimation of the end-to-end
performance over the dual-hop path $k$ using the \textit{selection
variable} $W_k(t)$, which is estimated separately by each relay
$R_k$ in the beginning of each slot from its channel state
information (CSI) \cite{1}. To facilitate channel state estimation
by the relays, $S$ and $D$ previously exchange short control
packets. In the beginning of slot $t$, the ``best" relay $b$ is
selected in a distributed manner by using the \textit{selection
policy}:
\begin{equation}
b = \mathop{\arg\max} \limits_{{}^{1\leq k\leq M}} \{W_k(t)\}.
\end{equation}
\subsubsection{Decode-and-forward relaying}
In the beginning of slot $t$, each DF relay estimates the
selection variable
\begin{equation}\label{1}
W_k^{DF}(t) = \min\{\alpha_{Sk}(t),\alpha_{kD}(t)\} \,,
\end{equation}
which actually evaluates the minimal instantaneous received SNR
among the two hops, $\min\{\gamma_{Sk}(t), \gamma_{kD}(t)\}$ .

\subsubsection{Amplify-and-forward relaying}
We assume fixed gain AF relays that amplify the received signal
from the first hop by $\sqrt{P_T /(P_T \Omega_{Sk} +N_0)}$ and
forward it to $D$ over the second hop \cite{9}. In the beginning
of slot $t$, each AF relay estimates the selection variable
\begin{equation}\label{2}
W_{k}^{AF}(t)=\frac{\alpha_{Sk}(t)\alpha_{kD}(t)}{\sqrt{C_k+\alpha_{kD}^2(t)}}
\,,
\end{equation}
where $C_k = \Omega_{Sk}+N_0/P_T$. Actually, each AF relay
evaluates the dual-hop SNR that is relayed over $R_k$ and received
at $D$. Note that fixed-gain (e.g., semi-blind) AF relays have
considerably simpler but yet comparably close performance to that
of the variable gain AF relays \cite{9}. \vspace{-0.2cm}
\subsection{Stations' mobility }
We consider 2-dimensional isotropic scattering around source $S$,
relays $R_k$ and destination $D$, all of which are assumed to be
mobile and have no line-of-sight with other stations. Thus, each
$S\rightarrow R_k$ ($R_k\rightarrow D$) hop behaves as a
mobile-to-mobile Rayleigh channel. Its channel gain,
$\alpha_{Sk}(t)$ ($\alpha_{kD}(t))$, is a time-correlated Rayleigh
random process with known statistical properties (e.g., the
Doppler spectrum) \cite{7}. If a station at one end of a hop is
fixed, the mobile-to-mobile Rayleigh-fading hop is transformed
into the ``classic" fixed-to-mobile Rayleigh-fading hop \cite{6}.
The time derivative $\dot\alpha_{Sk}$ ($\dot\alpha_{kD}$) is
independent from the channel gain $\alpha_{Sk}$ ($\alpha_{kD}$),
and follows the Gaussian probability distribution function (PDF)
with zero mean and variance [\ref{ref7}, Eq. (A5)] [\ref{ref10},
Eq. (39)]
\begin{eqnarray}
\sigma_{\dot\alpha_{Sk}}^2&=&\pi^2\Omega_{Sk}(f_{mS}^2+f_{mk}^2)\,,\label{3a}\\
\sigma_{\dot\alpha_{kD}}^2&=&\pi^2\Omega_{kD}(f_{mk}^2+f_{mD}^2)\,.\label{3b}
\end{eqnarray}
In (\ref{3a})-(\ref{3b}), $f_{mS}$, $f_{mD}$ and $f_{mk}$ denote
the maximum Doppler rates of  $S$, $D$ and relay $R_k$ $(1\leq k
\leq M )$, respectively. The Doppler rate (and consequently the
AOR) is expressed in the unit $slot^{-1}$. Expressing the slot
duration in seconds, both the Doppler rate (now becoming the
Doppler frequency) and the AOR are expressed in Hz. \vspace{-3mm}
\subsection{Capacity outage events}
In a given slot $t$, the opportunistic relaying system experiences
a capacity outage event when the mutual information of the
dual-hop path over the ``best" relay $b$ drops below some
predefined spectral efficiency $R$ \cite{2},
\begin{equation}\label{4}
I(t)=\frac{1}{2}\log_2\left( 1+\frac{P_T}{N_0}(W_{\max}(t))^2
\right)\leq R \,,
\end{equation}
where \vspace{-3mm}
\begin{equation}\label{5}
W_{\max}(t)=\max\{W_{1}(t),W_{2}(t),\dots, W_{M}(t) \} \,.
\end{equation}
Thus, the resulting time-varying capacity $I(t)$ suffers from the
random occurrence of capacity outage events, during which the
channel is unable to support the specified $R$. Transforming
(\ref{4}), the capacity outage event at slot $t$ occurs if
\begin{equation}\label{6}
W_{\max}(t) \leq Z \,,
\end{equation}
where the outage threshold is $Z=\sqrt{(2^{2R}-1)/(P_T/N_0)}$.
Note, $Z$ can be varied by varying $R$ (as in \cite{5b}) or
$P_T/N_0$ (as in this work), but the functional dependencies of
the increasing $R$ or the decreasing $P_T/N_0$ (in dB) have almost
same shapes. \vspace{-0.5cm}
\section{Average Outage Rates and Outage Durations}
\subsection{General expression}
Using a similar approach to that presented in \cite{8} for
deriving level crossing rates (LCR) of ``classic" selection
diversity systems, the joint PDF of $W_{\max}$ and $\dot W_{\max}$
is expressed as
\begin{equation}\label{7}
f_{W_{\max} \dot W_{\max}}(w,\dot w)=\sum_{k=1}^M  f_{W_{k} \dot
W_{k}}(w,\dot w) P_k(w)
\end{equation}
where $f_{W_{k} \dot W_{k}}(w,\dot w)$ denotes the joint PDF of
selection variable $W_k(t)$ and its time derivative $\dot W_k(t)$.
Assuming $R_k$ is selected ``best" relay, $P_k(w)$ denotes the
conditional probability that the selection variable $W_k$ of $R_k$
drops below $w$,
\begin{equation}\label{8}
P_k(w)= {\rm {Pr}} \{W_k\leq w\big | W_k \textrm{ is
max}\}=\prod_{i=1,i\neq k}^M F_{W_i}(w) \,,
\end{equation}
where $F_{W_i}(w)=\textrm{Pr}\{W_i\leq w\}$  denotes the
cumulative distribution function (CDF) of $W_i$. The AOR is
evaluated based on the standard LCR definition [\ref{ref6},
Chapter 1], yielding
\begin{eqnarray}\label{9}
N(Z) \stackrel{\rm def}{=} \int_0^\infty \dot w f_{W_{\max} \dot W_{\max}}(Z,\dot w) d\dot w \qquad \qquad \qquad \qquad \nonumber\\
=\sum_{k=1}^M P_k(Z) \int_0^\infty \dot w f_{W_{k},\dot
W_{k}}(Z,\dot w) =\sum_{k=1}^M P_k(Z) N_k(Z),
\end{eqnarray}
where $N_k(Z)$ denotes the AOR of the dual-hop path over the relay
$R_k$. The AOD is then given by
\begin{eqnarray}\label{10}
T(Z) \stackrel{\rm def}{=} \frac{\textrm{Pr}\{W_{\max}\leq Z\}}{N(Z)} \qquad \qquad \qquad \qquad \qquad \qquad \nonumber\\
= N(Z)^{-1}\prod_{k=1}^M F_{W_k}(Z) = \left
(\sum_{k=1}^M\frac{N_k(Z)}{F_{W_k}(Z)}\right)^{-1} .
\end{eqnarray}
\vspace{-3mm}
\subsection{Decode-and-forward relaying}
We now focus on the random process $W_k^{DF}(t)$, defined by
(\ref{1}). Communication through the relay $R_k$ falls in outage
if either one of the two hops fail. Thus, the PDF of $W_k^{DF}(t)$
is given by
\begin{equation}\label{11}
f_{W_k}(w)=f_{\alpha_{Sk}}(w) \textrm{Pr}\{ \alpha_{kD}>w\}+
f_{\alpha_{kD}}(w) \textrm{Pr}\{ \alpha_{Sk}>w\}.
\end{equation}
Since both hops follow the Rayleigh PDF, $W_k^{DF}$  is also
determined to follow the Rayleigh PDF,
\begin{equation}\label{12}
f_{W_k}(w)=2\Lambda_k w e^{-\Lambda_k w^2} \,,
\end{equation}
with $\Lambda_k = \Omega_{Sk}^{-1}+\Omega_{kD}^{-1}$, and the
respective CDF $F_{W_k}(w) = 1-\exp(-\Lambda_k w^2)$. Using
\begin{equation}\label{13}
\dot W_k=\left\{
\begin{array}{c}
\dot \alpha_{Sk}, \quad \alpha_{Sk} \leq \alpha_{kD}\\
\dot \alpha_{kD}, \quad \alpha_{kD}< \alpha_{Sk}
\end{array}
\right.
\end{equation}
and the independence of the channel gains and their respective
time derivatives, the required joint PDF is found as
\begin{eqnarray}\label{14}
f_{W_k \dot W_k}(w,\dot w) = f_{\dot \alpha_{Sk}}(\dot w)
f_{\alpha_{Sk}}(w) \textrm{Pr}\{ \alpha_{kD}>w\} \nonumber\\
+ f_{\dot \alpha_{kD}}(\dot w)  f_{\alpha_{kD}}(w)
\textrm{Pr}\{\alpha_{Sk}>w\} .
\end{eqnarray}
In Rayleigh fading, (\ref{14}) specializes to
\begin{equation}\label{15}
f_{W_k \dot W_k}(w,\dot w)=f_{W_k}(w) f_{\dot W_k}(\dot w) ,
\end{equation}
thus rendering $W_k$ and $\dot W_k$ as independent RVs, where
$f_{W_k}(w)$  is given by (\ref{12}) and
\begin{equation}\label{16}
f_{\dot W_k}(\dot w)=\frac{\Omega_{kD}}{\Omega_{Sk}+\Omega_{kD}}
f_{\dot \alpha_{Sk}}(\dot w) +
\frac{\Omega_{Sk}}{\Omega_{Sk}+\Omega_{kD}} f_{\dot
\alpha_{kD}}(\dot w)
\end{equation}
with $f_{\dot \alpha_{Sk}}(\cdot)$  and $f_{\dot \alpha_{kD}}
(\cdot)$ denoting zero mean Gaussian PDFs with variances
$\sigma_{\dot\alpha_{Sk}}^2$ (\ref{3a}) and
$\sigma_{\dot\alpha_{kD}}^2$ (\ref{3b}), respectively. Thus, the
AOR of dual-hop path over $R_k$ is obtained as

\begin{equation}\label{17}
N_k(Z)=\frac{\Omega_{kD} \sigma_{\dot\alpha_{Sk}} + \Omega_{Sk}
\sigma_{\dot\alpha_{kD}}
}{\Omega_{Sk}+\Omega_{kD}}\frac{f_{W_k}(Z)}{\sqrt{2\pi}}
\end{equation}
with $f_{W_k}(\cdot)$  given by (\ref{12}). Inserting (\ref{17})
into (\ref{9}) and (\ref{10}), we obtain the AOR and the AOD of DF
relaying system. \vspace{-3mm}

\begin{figure}
\centering
\includegraphics[width=3.3in]{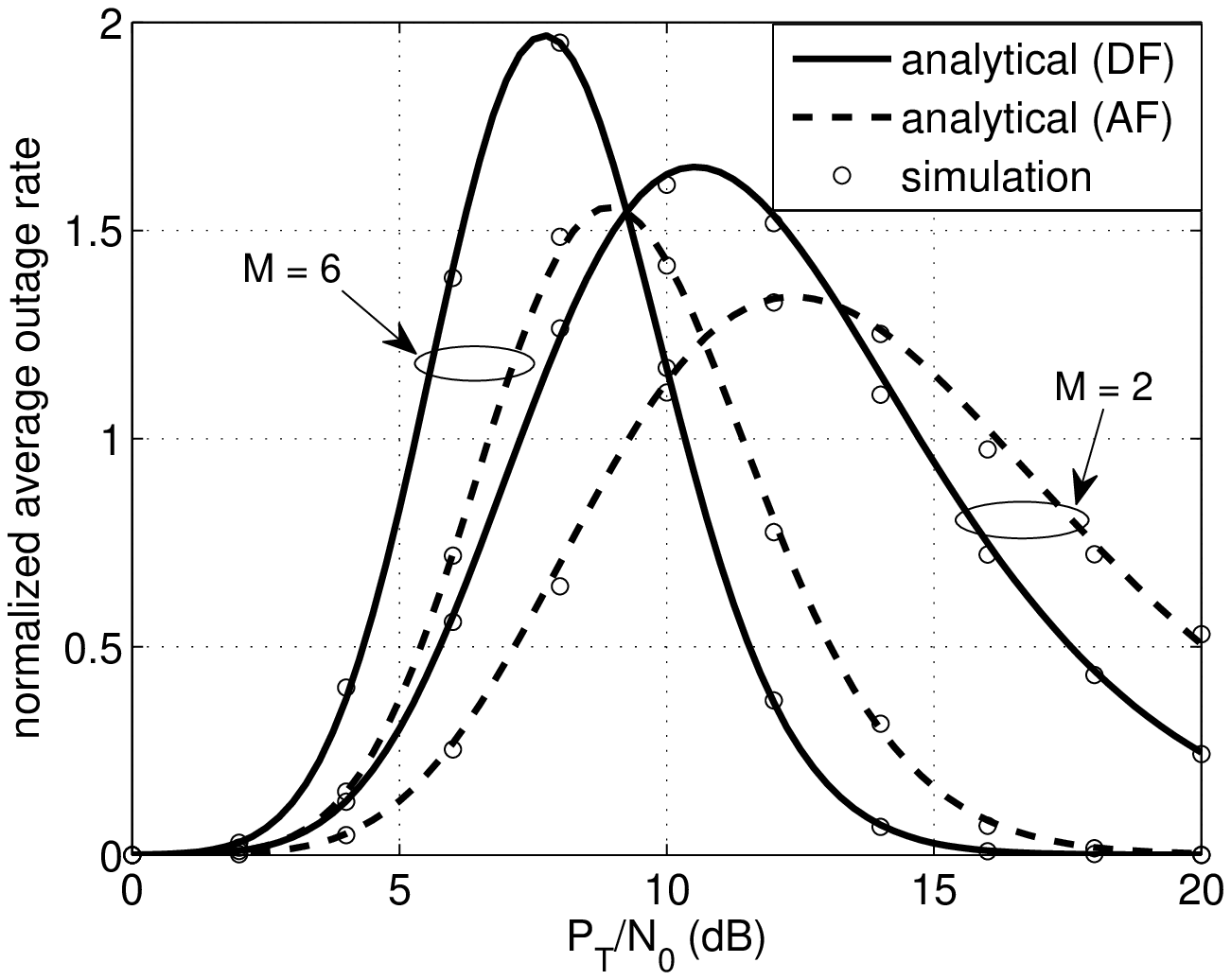}
\caption{Normalized Average Outage Rate in DF and AF opportunistic
relaying systems} \label{fig_1} \vspace{-0.4cm}
\end{figure}

\subsection {Amplify-and-forward relaying }
The exact expressions for the CDF and the AOR of the random
process $W_k^{AF}(t)$, defined by (\ref{2}), are respectively
given by [\ref{ref9}, Eq. (9)] and [\ref{ref10}, Eq. (19)], as
\begin{eqnarray}
F_{W_k}(Z) = 1-2Z\sqrt{\frac{C_k}{\Omega_{Sk} \Omega_{kD}}}
\exp\left(-\frac{Z^2}{\Omega_{Sk}}\right) \qquad \nonumber\\
 \times \, K_1\left( 2Z\sqrt{\frac{C_k}{\Omega_{Sk} \Omega_{kD}}}
 \right) \,, \label{18} \\
N_k(Z)=\sqrt{\frac{2}{\pi}}\frac{2Z\exp(-Z^2/\Omega_{Sk})}{\Omega_{Sk} \Omega_{kD}} \qquad \qquad \qquad \quad \nonumber\\
 \times\int_0^\infty   \sqrt{\sigma_{\dot\alpha_{Sk}}^2  (y^2+C_k) + \frac{1}{y^4} \sigma_{\dot\alpha_{kD}}^2 C_k^2 Z^2 } \quad \nonumber\\
 \times \exp\left(-\frac{\Omega_{Sk} y^4 +C_k \Omega_{kD} Z^2}{y^2 \Omega_{Sk} \Omega_{kD}}\right) dy \,, \label{19}
\end{eqnarray}
where $K_1(\cdot)$ is the first-order modified Bessel function of
the second kind. Note that (\ref{19}) can be efficiently and
accurately evaluated by applying the Gauss-Hermite quadrature rule
[\ref{ref11}, Eq. (25.4.46)]. Combining (\ref{18}) and (\ref{19})
into (\ref{9}) and (\ref{10}), we obtain the AOR and the AOD of AF
relaying system.

Note, if source $S$ and destination $D$ are fixed, the approach
presented in this section can be applied to derive analogous
analytic expressions for AORs and AODs for more general fading
channels (such as, Rice and Nakagami-$m$ models), because such
fixed-to-mobile hops have known second-order statistical
properties. \vspace{-3mm}

\begin{figure}
\centering
\includegraphics[width=3.3in]{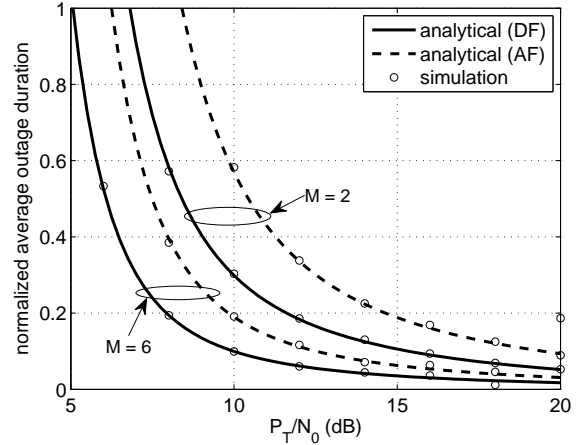}
\caption{Normalized Average Outage Duration in DF and AF
opportunistic relaying systems} \label{fig_2}
\vspace{-0.4cm}
\end{figure}

\section{Numerical and simulation examples }
In this section, we present illustrative examples for the
normalized AOR (Fig. 1) an the normalized AOD (Fig. 2) in function
of $P_T/N_0$ of opportunistic relaying system employing either DF
or AF relays. The source $S$ and the destination $D$ are fixed
($f_{mS} = f_{mD} = 0$), whereas all the relays are mobile and
introduce same maximum Doppler rates $f_{mk} = f_{m0}$, $1 \leq k
\leq M$. The AOR and the AOD are normalized with respect to the
Doppler rate $f_{m0}$ as $N/f_{m0}$ and $T \cdot f_{m0}$. The
average squared channel gains of all hops are equal to $0.5$
(i.e., $\Omega_{Sk} = \Omega_{kD} = 0.5$, $1 \leq k \leq M$), thus
rendering total available transmission power equal to $P_T$. The
spectral efficiency is set to $R = 1$ bps/Hz. The Monte Carlo
simulations clearly validate our derived analytical results.

\end{document}